\documentclass[nojss
]{jss}

\usepackage[utf8]{inputenc}

\providecommand{\tightlist}{%
  \setlength{\itemsep}{0pt}\setlength{\parskip}{0pt}}

\author{
Matthew Trupiano\\Rochester Institute of Technology
}
\title{The \proglang{R} Package \pkg{knnwtsim}: Nonparametric
Forecasting With a Tailored Similarity Measure}

\Plainauthor{Matthew Trupiano}
\Plaintitle{The R Package knnwtsim: Nonparametric Forecasting With a
Tailored Similarity Measure}
\Shorttitle{\pkg{knnwtsim}: Nonparametric Forecasting With a Tailored
Similarity Measure}

\Abstract{
The \proglang{R} package \pkg{knnwtsim} provides functions to implement
k nearest neighbors (KNN) forecasting using a similarity metric tailored
to the forecasting problem of predicting future observations of a
response series where recent observations, seasonal or periodic
patterns, and the values of one or more exogenous predictors all have
predictive value in forecasting new response points. This paper will
introduce the similarity measure of interest, and the functions in
\pkg{knnwtsim} used to calculate, tune, and ultimately utilize it in KNN
forecasting. This package may be of particular value in forecasting
problems where the functional relationships between response and
predictors are non-constant or piece-wise and thus can violate the
assumptions of popular alternatives. In addition both real world and
simulated time series datasets used in the development and testing of
this approach have been made available with the package.
}

\Plainkeywords{knn, time
series, forecasting, nonparametric, similarity, R}

\Address{
    Matthew Trupiano\\
    Rochester Institute of Technology\\
    School of Mathematical Sciences\\
  E-mail: \email{met2960@rit.edu}\\
  
  }

\usepackage{amsmath} \usepackage{framed} \usepackage{float} 

\begin{document}

\section{Introduction} \label{section:introduction}

The motivation in creating \pkg{knnwtsim} is to provide an additional
method to the space of forecasting univariate time series where all or
some combination of recent observations, seasonal or periodic patterns,
and the values of one or more exogenous predictors provide value in the
estimation of future observations of the series. \pkg{knnwtsim}
accomplishes this by providing functions to generate a weighted
similarity measure, referred to as \(S_w\) throughout this paper, which
has the ability to account for all three of these components and tune
their contribution to the overall similarity ultimately used in the
identification of nearest neighbors. Once the similarity matrix is
calculated the package provides functionality to use \(S_w\) in k
nearest neighbors (KNN) regression to forecast future points. This paper
will introduce the formulation of this measure and how KNN forecasting
is implemented in \pkg{knnwtsim}.

In terms of the current forecasting landscape, a popular and often
excellent alternative method used in this forecasting scenario is
regression with autoregressive integrated moving average (ARIMA) errors.
\begin{equation}
y_t = \mathbf{b}^{\top}\mathbf{x}_t + \eta_t
\label{eq:reg_with_arima}
\end{equation} Where \(y_t\) represents the value of the response series
of interest at time order \(t\), \(\mathbf{b}\) a vector of
coefficients, \(\mathbf{x}_t\) a vector of the values at time order
\(t\) of exogenous predictors, and \(\eta_t\) represents an ARIMA error,
see \citep[][Chapter~10.4]{fpp3}. In \proglang{R} \cite{R} this
technique is implemented in packages such as \pkg{forecast}
\cite{forecast} and \pkg{fable} \cite{fable}. In this paper an example
with simulated data of a situation where the use of KNN forecasting with
\pkg{knnwtsim} is better suited than regression with ARIMA errors is
explored in Section \ref{section:simulatedex}.

Also in \proglang{R} is an alternative package for KNN forecasting,
\pkg{tsfknn} \cite{tsfknn}. This package also provides functions for
univariate forecasting with KNN, however \pkg{knnwtsim} differs in
providing the functions to develop \(S_w\), see Equation \ref{eq:S_w} in
Section \ref{section:simform}, as well as providing a
\code{knn.forecast} function which can take any similarity matrix
regardless of the exact method used to create it. This makes
\pkg{knnwtsim} more tailored to situations where exogenous predictors
are involved.

\section {Similarity formulation} \label{section:simform}

The key to tailoring KNN to the specific time series problem at hand is
the similarity measure used to determine the neighborhood, formulated in
Equation \ref{eq:S_w}. \begin{equation}
S_w = \alpha S_t + \beta S_p + \gamma S_x
\label{eq:S_w}
\end{equation} \(S_t\), \(S_p\), and \(S_x\) are all separate component
similarity measures used to compare a given point \(y_t\) to other
previous points, \(y_i\) for \(i < t\), based on different aspects of
the time series of interest. \(S_t\) measures similarity in the sense of
time order, i.e., the closest prior point to \(y_t\) in terms of \(S_t\)
should be \(y_{t-1}\). If this metric was used alone to determine the
neighborhood in KNN regression, the estimate of \(y_t\) would simply be
a rolling average of the \(k\) most recent points prior to time \(t\).
\(S_p\) measures similarity in terms of periodicity, meaning if a time
series represented monthly data which typically follow a yearly cycle,
\(p\) would take a value between 1-12, and the points closest to \(y_t\)
should be the \(y_i\) where
\(i=\left(t-ap_{max}\right)=\left(t-a12\right)\). where \(a\) is an
integer greater than 1 that represents the number of years back in the
series. This means all prior points which fall in the same month of any
prior year would be considered equally similar by this component of
\(S_w\). Finally, to incorporate the information provided by exogenous
predictors which could be valuable in forecasting future points in the
series, there is \(S_x\). For a given point of interest \(y_t\) this
similarity metric compares a vector of predictors at time \(t\),
\(\mathbf{x}_t\), associated with \(y_t\), to each vector of predictors
\(\mathbf{x}_i\) associated with each prior point in the series,
\(y_i\).

\(\alpha\), \(\beta\), and \(\gamma\) represent the three weights that
can be tuned as hyperparameters. It is recommended to normalize these so
that they sum to 1, which allows for a rough physical interpretation of
these weights as measures of the relative importance of each of these
three aspects of the series targeted by the component similarity
measures in determining the future values of the series of interest.

\subsection{Implemented component similarity calculation methods} \label{section:simcalc}

In principle there is no strict requirement on the similarity metric
used to compute each of the component similarities which combine to form
\(S_w\). However here I detail the metrics implemented in the
\pkg{knnwtsim}. In this package each of the three similarity metrics:
\(S_t\), \(S_p\), and \(S_x\), are first calculated as a dissimilarity
metric: \(D_t\), \(D_p\), \(D_x\). These dissimilarities are then
transformed into the final similarity measures by \(\frac{1}{1+D_i}\)
for \(i \in t,p,x\), see \citep[][Chapter~10.2.1]{Bajorski}.

The measure implemented for \(D_t\) to calculate the dissimilarity
between a point \(y_t\) and another point \(y_i\), at time orders \(t\)
and \(i\) respectively is shown in Equation \ref{eq:D_t}.
\begin{equation}
D_t\left(t,i\right)=\left|t-i\right|
\label{eq:D_t}
\end{equation} Which is equivalent the standard Euclidean distance.

The metric used for dissimilarity in terms of periodicity, \(D_p\), is
formulated as follows. Between a point \(y_t\) and another point \(y_i\)
with respective periods \(p_t\) and \(p_i\), for example in a monthly
series taking on values between 1-12, we have Equation \ref{eq:D_p}.
\begin{equation}
D_p\left(p_t,p_i\right)=min\left[\left|p_t-p_i\right|\ ,\left|min\left(p_t,p_i\right)-p_{min}\right|+\left|p_{max}-max\left(p_t,p_i\right)\right|+1\right]
\label{eq:D_p}
\end{equation} Where \(p_{min}\) is the minimum value the period can
take on, and \(p_{max}\) the maximum, 1 and 12 respectively in the
monthly example. This formulation is based on the idea that generally
the periods at the very end and very beginning of a cycle should be
fairly similar. To clarify through an example, in a monthly cycle where
1 represents January and 12 represents December, these two months are
generally more similar to each-other than either are to July at period
7. Continuing with this monthly example, if we have three points which
occur in January, March, and November: \(p_i = 1\), \(p_j = 3\),
\(p_k = 11\) with \(p_{max} = 12\), and \(p_{min}\) is assumed to always
be 1, matching the behavior of \pkg{knnwtsim}. Then
\(D_p \left(p_i,p_j \right) = 2\), \(D_p \left(p_i,p_k\right) = 2\), and
\(D_p \left(p_j,p_k\right) = 4\).

For \(D_x\), the metric used to determine the dissimilarity between two
points \(y_t\) and \(y_i\) based on the values of associated exogenous
predictors \(\mathbf{x}_t\) and \(\mathbf{x}_i\), any commonly known and
appropriate distance metric taking the two vectors of predictors as
input would work. For example, in situations where predictors are
binary, such as holiday indicators, one could use \texttt{binary}
distance as implemented in the \proglang{R} package \pkg{stats} \cite{R}
function \texttt{dist}. In situations where predictors are numeric,
using Euclidean or Mahalanobis distance may be preferred, again see
\citep[][Chapter~10.2.1]{Bajorski} for additional details.

\section{KNN forecasting} \label{section:knnforecasting}

Once a final similarity matrix is calculated, not necessarily using the
measures listed above, that matrix can be used in KNN regression to
forecast future observations of a response series. The model for KNN
regression as used in \pkg{knnwtsim} can be formulated as shown in
Equation \ref{eq:KNN_model}, see
\citep[][Chapter~2.4]{Clarke+Fokoue+Zhang}. \begin{equation}
y_t = \frac{1}{k}\sum_{i\in K\left(y_t\right)} y_i + \epsilon_t
\label{eq:KNN_model}
\end{equation} Where \(k\) is the number of nearest neighbors
considered, and a hyperparameter which will generally need to be tuned.
\(K\left(y_t\right)\) is the ``neighborhood'' consisting of the \(k\)
observations of \(y_i\) with the highest similarity (or lowest
dissimilarity/distance) to \(y_t\) of all prior members of the time
series, meaning \(i\ <\ t\). \(\epsilon_t\) is an error term to capture
the remaining unexplained random component, which has not been required
to follow any particular distributional form.

Naturally we may want to forecast an arbitrarily long forecast horizon,
\(h\), where \(h\ \geq 1\). The estimate for any given future point is
shown below as Equation \ref{eq:KNN_est}. With \(i\ <\ t\) and \(t\)
being in this example the last observed actual observation.
\begin{equation}
{\hat{y}}_{t+h}=\frac{1}{k}\sum_{i\in K\left({\hat{y}}_{t+h}\right)} y_i
\label{eq:KNN_est}
\end{equation} To date in \pkg{knnwtsim} previously estimated points
have not been included as potential candidates for the neighborhood,
only actual observations. To elaborate with an example, if the latest
observation is at time \(t\) and one is interested in forecasting out to
time \(t+h\), with \(h>1\), then \({\hat{y}}_{t+1}\) would not be a
viable option to include in \(K\left({\hat{y}}_{t+h}\right)\) based on
the behavior programmed in the function \texttt{knn.forecast} in the
\pkg{knnwtsim} package.

\section[The knnwtsim Package]{The \pkg{knnwtsim} Package} \label{section:package}

This section provides an overview to the key functions within
\pkg{knnwtsim} a user may need when working through a forecasting
exercise with the package.

\subsection{Functions for similarity matrices}

Each of of the three component similarity measures used in generating
\(S_w\) has a corresponding function in \pkg{knnwtsim} to generate a
similairty matrix using that measure as defined in Section
\ref{section:simcalc}.

\begin{enumerate}
\def\labelenumi{\arabic{enumi}.}
\item
  \code{StMatrixCalc} produces a similarity matrix using \(S_t\) when
  provided with a vector of time orders corresponding to the values of
  the response series. The only argument is:

  \begin{itemize}
  \tightlist
  \item
    \code{v}: A numeric vector with the time order corresponding to each
    point in the response series.
  \end{itemize}
\item
  \code{SpMatrixCalc} produces a similarity matrix using \(S_p\) when
  provided a vector of seasonal periods corresponding to the values of
  the response series and the total number of periods in a full cycle,
  12 for monthly data as an example. The arguments are:

  \begin{itemize}
  \tightlist
  \item
    \code{v}: A numeric vector with the seasonal periods corresponding
    to each point in the response series.
  \item
    \code{nPeriods}: A numeric value representing the maximum value
    \code{v} can take on.
  \end{itemize}
\item
  \code{SxMatrixCalc} given a numeric vector or matrix of the values of
  exogenous predictors corresponding to the response series and a
  distance calculation method, \code{SxMatrixCalc} calls the \pkg{stats}
  package's \code{dist} function using the method provided and
  transforms the resulting matrix into a similarity measure which is the
  returned to the user. The arguments are:

  \begin{itemize}
  \tightlist
  \item
    \code{A}: A numeric vector or matrix where the columns represents
    exogenous predictor variables and the rows correspond to the points
    in the response series.
  \item
    \code{XdistMetric}: A string describing the method the \pkg{stats}
    package's \code{dist} function should use. This must be one of
    \code{"euclidean"}, \code{"maximum"}, \code{"manhattan"},
    \code{"canberra"}, \code{"binary"}, or \code{"minkowski"}.
  \end{itemize}
\end{enumerate}

All of the matrices returned by these functions will be of dimension
\(n~x~n\) where \(n\) is the length of the input vector or row count of
the input matrix in the case of \code{SxMatrixCalc}. The final of the
similarity matrix functions calls each of these components.

\begin{enumerate}
\def\labelenumi{\arabic{enumi}.}
\setcounter{enumi}{3}
\item
  \code{SwMatrixCalc} given all of the arguments to the component matrix
  functions plus a vector of weights, calls \code{StMatrixCalc},
  \code{SpMatrixCalc}, and \code{SxMatrixCalc} to generate each
  component matrix, then the first weight in the provided vector is
  multplied by the matrix using \(S_t\), the second weight is multiplied
  by the matrix generated using \(S_p\), and the third weight is
  multiplied by the matrix generated using \(S_x\). Finally, the three
  weighted component matrices are added to produce a single matrix which
  is returned. Each element of this matrix will then have been
  calculated as per Equation \ref{eq:S_w}. The arguments are:

  \begin{itemize}
  \tightlist
  \item
    \code{t.in}: A numeric vector with the time order corresponding to
    each point in the response series.
  \item
    \code{p.in}: A numeric vector with the seasonal periods
    corresponding to each point in the response series.
  \item
    \code{nPeriods.in}: A numeric value representing the maximum value
    \code{p.in} can take on.
  \item
    \code{X.in}: A numeric vector or matrix where the columns represents
    exogenous predictor variables and the rows correspond to the points
    in the response series.
  \item
    \code{XdistMetric.in}: A string describing the method the
    \pkg{stats} package's \code{dist} function should use. This must be
    one of \code{"euclidean"}, \code{"maximum"}, \code{"manhattan"},
    \code{"canberra"}, \code{"binary"}, or \code{"minkowski"}.
  \item
    \code{weights}: A numeric vector where first value represents weight
    for \(S_t\), second value the weight for \(S_p\), and the third
    value the weight for \(S_x\).
  \end{itemize}
\end{enumerate}

\subsection{Function for KNN forecasting}

The core user facing function for forecasting using KNN regression as
defined in Equation \ref{eq:KNN_est} is \code{knn.forecast}. This
function takes a similarity matrix as input, along with an index
corresponding to which time orders of the response series should be
estimated, the number of nearest neighbors, \(k\), to be used in
estimation, and the response series for which the forecast is being
generated. This function uses the input forecasting index to identify
the columns of the similarity matrix which should be sorted to identify
the nearest neighbors, and also removes the rows at this same index to
exclude observations in the forecast index from consideration as
neighbors. Naturally with this construction the index in terms of both
rows and columns of the similarity matrix should be aligned to the index
of response series. Next it applies the function \code{NNReg} to each
column of interest, which sorts the provided column by greatest to least
similarity, identifies the indices of the nearest neighbors, and returns
the mean of the points in the response series at the nearest neighbor
indices. The \code{knn.forecast} function returns this value provided by
\code{NNReg} for each point in the forecast index as a numeric vector.
This vector is the KNN regression forecast. The arguments are:

\begin{itemize}
\tightlist
\item
  \code{Sim.Mat.in}: A numeric matrix of similarities .
\item
  \code{f.index.in}: A numeric vector indicating the indices of
  \code{Sim.Mat.in} and \code{y.in} which correspond to the time order
  of the points to be forecast.
\item
  \code{k.in}: An integer value indicating the the number of nearest
  neighbors to be considered in forecasting.
\item
  \code{y.in}: A numeric vector of the response series to be forecast.
\end{itemize}

\subsection{Tuning the hyperparameters}

There is one function included with the package for the tuning of the
weights, \(\alpha\), \(\beta\), and \(\gamma\) used to create \(S_w\),
and the number of neighbors \(k\). This is
\code{knn.forecast.randomsearch.tuning} which takes as input the number
of hyperparameter sets to generate and test, the three component
matrices calculated using \(S_t\), \(S_p\), and \(S_x\), the response
series, an integer value for the number of points used to estimate and
evaluate a test forecast, a maximum value which \(k\) can take on, and
finally the number of points at the end of the series which should be
held out for later validation. In this case the arguments are listed
before the function's behavior to help clarify the explanation:

\begin{itemize}
\tightlist
\item
  \code{grid.len}: An integer value representing the number of
  hyperparameter sets to generate and test.
\item
  \code{St.in}: A numeric matrix of similarities, can be generated with
  \code{StMatrixCalc}.
\item
  \code{Sp.in}: A numeric matrix of similarities, can be generated with
  \code{SpMatrixCalc}.
\item
  \code{Sx.in}: A numeric matrix of similarities, can be generated with
  \code{SxMatrixCalc}..
\item
  \code{y.in}: A numeric vector of the response series to be forecast.
\item
  \code{test.h}: An integer value representing the number of points to
  include in the test forecast.
\item
  \code{max.k}: An integer value representing the maximum value of
  \(k\), \code{knn.forecast} should use, this will be set to
  \code{min(floor((length(y.in)) * .4), 50)} if \code{NA} is passed.
\item
  \code{val.holdout.len}: An integer value representing the number of
  observations at the end of the series to be removed before the test
  forecast points are decided if desired to leave a validation set after
  tuning.
\end{itemize}

The function starts by creating \code{grid.len} randomly generated
hyperparameter sets,\\
\(\{k, \alpha, \beta, \gamma\}\), where the weights \(\alpha\),
\(\beta\), and \(\gamma\) are normalized to sum to 1. In generating
values for \(k\) this function will draw a random integer between 1 and
the \code{max.k} argument supplied, or the default maximum \(k\) if
\code{max.k} is not supplied. An index indicating the points to estimate
and use as a test set is created by taking the final \code{test.h}
observations in the response series \code{y.in} remaining after the
final \code{val.holdout.len} have been removed. Then for each set of
hyperparameters a new similarity matrix using \(S_w\) is produced and
used in a call to \code{knn.forecast} to forecast the points in the test
index. Each estimated point for the response series at the test index is
compared to the actual value of \code{y.in} at the corresponding time
order using average percent error (APE). Defined for a given point
\(y_{t+h}\), \(h\) steps ahead of the latest time order where an
observation could be considered a valid neighbor, \(t\), below in
Equation \ref{eq:APE}.\\
\begin{equation}
APE_h = \left( y_{t+h}-{\hat{y}}_{t+h}\right) / y_{t+h}\left(100\right)
\label{eq:APE}
\end{equation} Where \(\hat{y}_{t+h}\) represents the estimate of
\(y_{t+h}\) from \code{knn.forecast}. Each of the \code{test.h} APE
values are then averaged to calculate the mean average percent error
(MAPE) over the test points for the forecast estimated using a given set
of hyperparameters. After testing all hyperparameter sets
\code{knn.forecast.randomsearch.tuning} finds the set with the lowest
associated MAPE value in forecasting the points of the response series
in the test set and deems them the ``optimum'' set to use. These top
performing weights are returned as part of a list including the
following components:

\begin{itemize}
\tightlist
\item
  \code{weight.opt}: A numeric vector of the three weights to generate
  \(S_w\) in \(\alpha\), \(\beta\), \(\gamma\) order which achieved the
  best performance in terms of MAPE.
\item
  \code{k.opt}: An integer value of the number of neighbors used in
  \code{knn.forecast} which achieved the best performance in terms of
  MAPE.
\item
  \code{Test.MAPE}: A numeric value of the MAPE result for the optimum
  hyperparamter set achieved on the test points.
\item
  \code{MAPE.all}: A numeric vector of all MAPE results, each
  observation corresponds to the row in \code{Grid} of the same index.
\item
  \code{Grid}: A data frame of all hyperparameter sets tested in the
  tuning.
\item
  \code{Sw.opt}: A numeric matrix of similarities calculated using
  \(S_w\), with the best performing set of hyperparameters.
\end{itemize}

\subsection{Data included with the package} \label{section:data}

In addition to the functions used to implement KNN regression
forecasting using \(S_w\), \pkg{knnwtsim} contains some time series data
sets which can be used to experiment with this package or any other time
series methodology one wishes. Three of the datasets are stored as data
frames of count data derived from files on the city open data portal for
Boston, MA, USA.

\begin{enumerate}
\def\labelenumi{\arabic{enumi}.}
\item
  \code{boston_fire_incidents_weekly}
\item
  \code{boston_fire_incidents_monthly}
\item
  \code{boston_911dispatch_weekly}
\end{enumerate}

Two are derived from the same source,
\code{boston_fire_incidents_weekly} which contains a weekly sum of fire
incidents in the City of Boston, and
\code{boston_fire_incidents_monthly} which contains a monthly sum of
fire incidents in Boston, both with data from the start of 2017 to the
end of July, 2021 \cite{boston+fire+data}. The final Boston based data
set is \code{boston_911dispatch_weekly} which contains the number of 911
dispatches by the City of Boston public safety agencies between November
1, 2010 and April 21, 2014 \cite{boston+911+data}. There are three
potential response series in this data set with columns for \code{BPD},
\code{EMS}, and \code{BFD}. Both \code{boston_fire_incidents_weekly} and
\code{boston_911dispatch_weekly} contain a series of holiday indicators
relevant to the state Massachusetts. Additionally, both
\code{boston_fire_incidents_weekly} and
\code{boston_fire_incidents_monthly} contain an indicator for the weeks
which occurred during a COVID-19 state of emergency in the state.
Finally, as well as holiday indicators \code{boston_911dispatch_weekly}
contains a series of month indicators as well.

In addition to the data frames based on Boston's open data, the package
contains simulated data. This data is stored as
\code{simulation_master_list}, which is a list of 20 sub-lists each
containing four categories of simulated time series and 31 total
elements which also contain the randomly drawn components used to create
the four series. The only difference between the four time series in a
given sub-list is the relationship between response and predictors, and
that one category, \code{series.mvnormx}, uses a different set of
predictors than the other three. Here the simulation procedure is
described. Unless otherwise specified all distribution based sampling is
performed using \pkg{stats}'s functions: \texttt{rnorm} for univariate
normal distributions, \texttt{rpois} for Poisson distributions, and
\texttt{runif} for continuous uniform distributions.

The first of these four categories of response series,
\code{series.mvnormx}, was formulated as follows in Equation
\ref{eq:Sim_data_mvn}. \begin{equation}
y_t=c+\mathbf{b}^{\top}\mathbf{x}_t+\beta_{sin}sin\left(\frac{2\pi t}{s}\right)+\beta_{cos}cos\left(\frac{2\pi t}{s}\right)+\eta_t+\epsilon_t
\label{eq:Sim_data_mvn}
\end{equation} Breaking down the components we have:

\begin{itemize}
 \item $c$ : This term represents and randomly chosen constant from a continuous uniform distribution over the range $\left[-20 , 20 \right]$
 \item $\mathbf{x}_t$ : A $d~x~1$ vector corresponding the $t^{th}$ row in a matrix $\mathbf{X}$ with $d$ columns, representing exogenous predictors. These observations are drawn from a multivariate normal distribution using the \pkg{MASS} package's \code{mvrnorm} function \cite{MASS} with mean vector, $\boldsymbol{\mu}_\mathbf{x}=0$, and variance-covariance matrix $\mathbf{\Sigma}_\mathbf{x}$, which for each simulated series is a randomly generated correlation matrix using the \pkg{clusterGeneration} package's \code{rcorrmatrix} function \cite{clusterGeneration}. 
 \item $\mathbf{b}$ : Is a $d~x~1$ vector of coefficients where each of the $d$ elements are randomly pulled from a continuous uniform distribution over the range $\left[-5,5\right]$
 \item $\beta_{sin}sin\left(\frac{2\pi t}{s}\right)+\beta_{cos}cos\left(\frac{2\pi t}{s}\right)$ : Is a combination of terms to simulate a constant seasonal effect. $s$ represents the periodicity of the series, randomly sampled from the set $\left[4,7,12\right]$.  $\beta_{sin}$ and $\beta_{cos}$ are both drawn from the same continuous uniform distribution as $\mathbf{b}$. 
 \item $\eta_t$: Represents a simulated ARIMA$\left(p,d,q\right)$ term meant to potentially induce some serial correlation into the simulated series. The orders $p$, $d$, $q$ are sampled from the set $\left[0,1,2\right]$. Orders beyond 2 were not considered as the constraints for the coefficient vectors $\Phi$ and $\Theta$ of AR$\left(p\right)$ and MA$\left(q\right)$ processes with $p$ or $q\ >\ 2$ become much more complex. Both the $p$ coefficients, $\Phi$,  for the AR$\left(p\right)$ process and the $q$ coefficients, $\Theta$, for the MA$\left(q\right)$ process are generated randomly (when required). For processes which are AR$\left(1\right)$ or MA$\left(1\right)$ the single coefficient is drawn from a continuous uniform distribution over the range $\left[-1,1\right]$. For  AR$\left(2\right)$ and MA$\left(2\right)$ there are more complex constraints and so the two necessary coefficients are drawn from the same continuous uniform distribution on the range $\left[-1,1\right]$ until the set of coefficients meets the constraints for that process.
 
 \item $\epsilon_t$: Is a term to add additional random noise, it is drawn from either a $N\left(\mu_\epsilon=0,\sigma_\epsilon^2\right)$ or $Poisson\left(\lambda_\epsilon\right)$ distribution depending on the result of a single Bernoulli trial with equal probability to either type of noise, where the parameters $\sigma_\epsilon$ and $\lambda_\epsilon$ are sampled from a continuous uniform distribution on the range $\left[.1,\ 20\right]$. The reason for allowing the inclusion of a $Poisson\left(\lambda_\epsilon\right)$ error term was to induce some right skewed behavior into the series, which is common in many real-world series.
\end{itemize}

The other three categories are built in similar fashion,
\code{series.lin.to.sqrt.x}, \newline \code{series.lin.coef.chng.x}, and
\code{series.quad.to.cubic.x} are described by the following more
general Equation \ref{eq:Sim_data_piecewise}. In these series all
components which are common with those shown in Equation
\ref{eq:Sim_data_mvn} remain the same. \begin{equation}
y_t=c+f\left(x_t\right)+\beta_{sin}sin\left(\frac{2\pi t}{s}\right)+\beta_{cos}cos\left(\frac{2\pi t}{s}\right)+\eta_t+\epsilon_t
\label{eq:Sim_data_piecewise}
\end{equation} However, instead of a set of predictors with a simple
linear relationship to the response, we have \(f\left(x\right)\) which
describes some piece-wise functional relationship between the response
\(y\) and a single predictor \(x\). In these series realizations of the
single predictor \(x\) are drawn from a \(N(\mu_x,\sigma_x)\)
distribution where \(\mu_x\) and \(\sigma_x\) are both drawn from
continuous uniform distributions over the ranges \(\left[-5,5\right]\)
and \(\left[.001,10\right]\) respectively. For the \(f\left(x\right)\)
component of \code{series.lin.to.sqrt.x} we have Equation
\ref{eq:lin_to_sqrt} where \(m\) is a scalar coefficient simulated from
a continuous uniform distribution over the range
\(\left[-5 , 5 \right]\), and \(bp\) is also some scalar numeric value
simulated from a continuous uniform distribution on the range
\(\left[\mu_x - \sigma_x , \mu_x + \sigma_x \right]\), however for the
\code{series.lin.to.sqrt.x} series specifically the minimum \(bp\) is
.001 to ensure the square root function is defined. \begin{equation}
f(x) = \begin{cases} 
      mx & x < bp \\
      \sqrt(x) &  x \geq bp 
   \end{cases}
\label{eq:lin_to_sqrt}
\end{equation} Moving to the \code{series.lin.coef.chng.x} category of
series, the \(f(x)\) is component is defined as Equation
\ref{eq:lin_coef_chng}, where \(m_1\) is the same coefficient as \(m\)
in Equation \ref{eq:lin_to_sqrt}, and \(m_2\) is a separately drawn
coeficient also from a continuous uniform distribution over the range
\(\left[-5 , 5 \right]\). \begin{equation}
f(x) = \begin{cases} 
      m_1 x & x < bp \\
      m_2 x &  x \geq bp 
   \end{cases}
\label{eq:lin_coef_chng}
\end{equation} Finally the last category of time series in
\code{simulation_master_list} are those stored as
\code{series.quad.to.cubic.x}. Which have a \(f(x)\) component based on
Equation \ref{eq:quad_to_cubic}, where \(m\) is again the same
coeficient from Equation \ref{eq:lin_to_sqrt}. \begin{equation}
f(x) = \begin{cases} 
      m x^2 & x < bp \\
      -m x^3 &  x \geq bp 
   \end{cases}
\label{eq:quad_to_cubic}
\end{equation}

\section{Real world data example} \label{section:realworldex}

In order to demonstrate this method alongside regression with ARIMA
errors on a ``real-world'' example, we can use the
\texttt{boston\_fire\_incidents\_weekly} data set included with
\pkg{knnwtsim}. This data set contains 239 observations of the weekly
count of fire incidents in the City of Boston, MA, USA, as well as a
series of holiday indicators relevant to the U.S. state Massachusetts,
and an indicator for the weeks which occurred during a COVID-19 state of
emergency in the state. As mentioned in Section \ref{section:data} this
data set was derived from raw incident data sourced from Boston's open
data portal. The data is accessed as follows, and the head of the data
frame's first 5 columns are printed.

\begin{CodeChunk}
\begin{CodeInput}
R> data("boston_fire_incidents_weekly")
R> head(boston_fire_incidents_weekly[, 1:5])
\end{CodeInput}
\begin{CodeOutput}
# A tibble: 6 x 5
  week       incidents new.years.ind christmas.ind thanksgiving.ind
  <date>         <int>         <dbl>         <dbl>            <dbl>
1 2017-01-01       739             1             0                0
2 2017-01-08       874             0             0                0
3 2017-01-15       772             0             0                0
4 2017-01-22       867             0             0                0
5 2017-01-29       778             0             0                0
6 2017-02-05       834             0             0                0
\end{CodeOutput}
\end{CodeChunk}

Once the data is loaded it can be used to generate the necessary
information for forecasting with \pkg{knnwtsim}. A vector of time
orders, \texttt{df\$t}, and a vector of periods, \texttt{df\$p}, as well
as a variable for the maximum period value, \texttt{p.max}, are created
for use in \texttt{StMatrixCalc} and \texttt{SpMatrixCalc}. This data
provides a good opportunity to point out the seasonal level used does
not necessarily need to be on the same grain as the response data. In
this case we use monthly seasonality in the construction of the
similarity matrix using \(S_p\) while the response observations are on a
weekly level. Then the response series is captured in the vector
\texttt{y}, and the matrix of exogenous predictors for use in
\texttt{SxMatrixCalc} is stored in \texttt{X}. A separate version of the
response series using the \texttt{ts} class is captured as \texttt{y.ts}
for use in the \pkg{forecast} package's function \texttt{auto.arima}
which willl be used later for comparison.

\begin{CodeChunk}
\begin{CodeInput}
R> df <- boston_fire_incidents_weekly
R> df$t <- c(1:nrow(df))
R> p.max <- 12
R> df$p <- lubridate::month(df$week)
R> y <- df$incidents
R> Xcols <- names(boston_fire_incidents_weekly[, 
+   3:ncol(boston_fire_incidents_weekly)]) 
R> X <- as.matrix(df[, names(df) %in% Xcols])
R> y.ts <-ts(df$incidents, frequency = 52)
\end{CodeInput}
\end{CodeChunk}

Next the time frames to be used for tuning and validation are
established. The weights \(\alpha\), \(\beta\), and \(\gamma\), as well
as the number of neighbors \(k\) will be tuned with the latest 52 weeks
data, \texttt{test.len}, after the 26 most recent points,
\texttt{val.len}, are removed.

\begin{CodeChunk}
\begin{CodeInput}
R> test.len <- 52
R> val.len <- 26
R> n <- nrow(df)
R> val.index <- as.vector(c((n -  val.len + 1):(n)))
\end{CodeInput}
\end{CodeChunk}

At this stage each of the component similarity matrices used in \(S_w\)
are calculated using \texttt{StMatrixCalc}, \texttt{SpMatrixCalc}, and
\texttt{SxMatrixCalc}. These matrices are then provided as input to the
\texttt{knn.forecast.randomsearch.tuning} function, along with the
number of randomly generated hyperparameter sets to be tested, the
response series, the test horizon to be used to assess forecast
accuracy, and the number of points which should be held out for
validation, using the \texttt{grid.len}, \texttt{y.in}, \texttt{test.h},
and \texttt{val.holdout.len} arguments respectively. After the tuning is
complete the similarity matrix using \(S_w\) generated with the
hyperparameter set which achieved the lowest MAPE result over the points
in the test set is returned as part of list and can be accessed for
later forecasting, as are the hyperparameters of this `optimum' set. The
hyperparameter set of interest is printed below.

\begin{CodeChunk}
\begin{CodeInput}
R> St.ts <- knnwtsim::StMatrixCalc(v = df$t)
R> Sp.ts <- knnwtsim::SpMatrixCalc(v = df$p, nPeriods = p.max)
R> Sx.ts <- knnwtsim::SxMatrixCalc(A = X, XdistMetric = 'binary')
R> set.seed(10)
R> tuning.list <- knnwtsim::knn.forecast.randomsearch.tuning(
+   grid.len = 10 ** 4,
+   St.in = St.ts,
+   Sp.in = Sp.ts,
+   Sx.in = Sx.ts,
+   y.in = y,
+   test.h = test.len,
+   val.holdout.len = val.len)
R> Sw.ts <- tuning.list$Sw.opt 
R> k.ts <- tuning.list$k.opt
R> weights.ts <- tuning.list$weight.opt
\end{CodeInput}
\end{CodeChunk}

\begin{CodeChunk}
\begin{CodeOutput}

 Tuned k 
\end{CodeOutput}
\begin{CodeOutput}
[1] 7
\end{CodeOutput}
\begin{CodeOutput}

 Tuned Weights: alpha, beta, gamma 
\end{CodeOutput}
\begin{CodeOutput}
[1] 0.717238399 0.009966304 0.272795297
\end{CodeOutput}
\end{CodeChunk}

With the final tuned similarity matrix and value of \(k\), the
\texttt{knn.forecast} function can be called to generate a forecast for
the points of the response series in the validation set, the last 26
weekly observations, and print the result.

\begin{CodeChunk}
\begin{CodeInput}
R> final.forecast <- knnwtsim::knn.forecast(
+   Sim.Mat.in = Sw.ts,
+   f.index.in = val.index,
+   k.in = k.ts,
+   y.in = y)
R> final.forecast
\end{CodeInput}
\begin{CodeOutput}
      214       215       216       217       218       219       220 
 798.4286  798.4286  778.0000  797.0000  797.0000  797.0000  785.0000 
      221       222       223       224       225       226       227 
 797.0000  797.0000  797.0000  797.0000  779.2857  797.0000  797.0000 
      228       229       230       231       232       233       234 
 797.0000  797.0000  797.0000  817.0000  797.0000  826.7143  991.0000 
      235       236       237       238       239 
 991.0000  903.2857 1050.1429 1050.1429 1050.1429 
\end{CodeOutput}
\end{CodeChunk}

We then produce a baseline forecast for comparison over the same time
frame using the \code{auto.arima} function from the \pkg{forecast}
package. First fitting a model, \code{arima.model}, using the
\code{y.arima.train} and \code{X.arima.train} objects which remove the
validation points for later forecasting, and printing the model
coefficients for reference. Forecasting on the validation data is then
performed using with the \code{forecast} function and the result is
stored in the variable \code{arima.forecast}.

\begin{CodeChunk}
\begin{CodeInput}
R> y.arima.train <- y.ts[-(val.index)]
R> X.arima.train <- as.matrix(
+   X[-(val.index), ],
+   ncol = ncol(X),
+   nrow = nrow(y.arima.train))
R> arima.model <- forecast::auto.arima(
+   y.arima.train,
+   xreg = X.arima.train,
+   allowdrift = TRUE)
R> summary(arima.model)$coef
\end{CodeInput}
\begin{CodeOutput}
                   ar1              intercept          new.years.ind 
             0.5717191            938.4408148             15.6140738 
         christmas.ind       thanksgiving.ind           veterans.ind 
          -123.9662197            -15.8939117            -10.7502893 
indigenous.peoples.ind              labor.ind              july4.ind 
            41.8251938            -15.5094518             34.1769499 
        juneteenth.ind           memorial.ind           patriots.ind 
           116.6317132             -6.0453609            -62.7740881 
       st.patricks.ind         presidents.ind                mlk.ind 
             0.3174248           -113.9509642              6.3251394 
         covid.soe.ind 
           -90.4243520 
\end{CodeOutput}
\begin{CodeInput}
R> y.arima.val <- y.ts[(val.index)]
R> X.arima.val <- as.matrix(
+   X[(val.index), ], 
+   ncol = ncol(X),
+   nrow = nrow(y.arima.val))
R> arima.forecast <- forecast::forecast(
+   arima.model,
+   h = val.len,
+   xreg = X.arima.val)$mean
\end{CodeInput}
\end{CodeChunk}

With the baseline forecast calculated the performance metrics for both
forecasts are assessed using MAPE, the mean of the APE values for each
point as calculated by Equation \ref{eq:APE}. In addition we can combine
the forecasts from the two different methods into what we will call the
\code{ensemble.forecast}, and assess the performance achieved from
combining both forecast results.

\begin{CodeChunk}
\begin{CodeInput}
R> final.actuals <- y[val.index]
R> knn.APE <- abs((final.actuals - final.forecast ) / final.actuals) * 100
R> knn.MAPE <- mean(knn.APE)
R> arima.APE <- abs((final.actuals - arima.forecast) / final.actuals) * 100
R> arima.MAPE <- mean(arima.APE)
R> ensemble.forecast <- (arima.forecast + final.forecast) / 2
R> ens.APE <- abs((final.actuals - ensemble.forecast) / final.actuals) * 100
R> ens.MAPE <- mean(ens.APE)
\end{CodeInput}
\end{CodeChunk}

The results as shown in Table \ref{fire-mape-table} indicate that the
ensemble forecast was the most accurate in terms of MAPE over the
forecast horizon at \(7.96\%\), followed by the forecast from
\code{knn.forecast} at \(8.14\%\), and finally the \code{auto.arima}
forecast at \(8.60\%\). Visually this comparison can be viewed as Figure
\ref{fig:fire-incidents-plot}, where the series after week 190 is
plotted and lines corresponding to the forecasts of the final 26 points
from \code{knn.forecast}, \code{auto.arima}, and the ensemble of the two
are plotted alongside their actual observations.

\begin{table}[t!]
\centering
\begin{tabular}{rrrr}
  \hline
 & KNN regression & Regression with ARIMA errors & Ensemble \\ 
  \hline
1 & 8.14 & 8.60 & 7.96 \\ 
   \hline
\end{tabular}
\caption{Fire incident MAPE results by forecast method.} 
\label{fire-mape-table}
\end{table}

\begin{CodeChunk}
\begin{figure}[t!]

{\centering \includegraphics{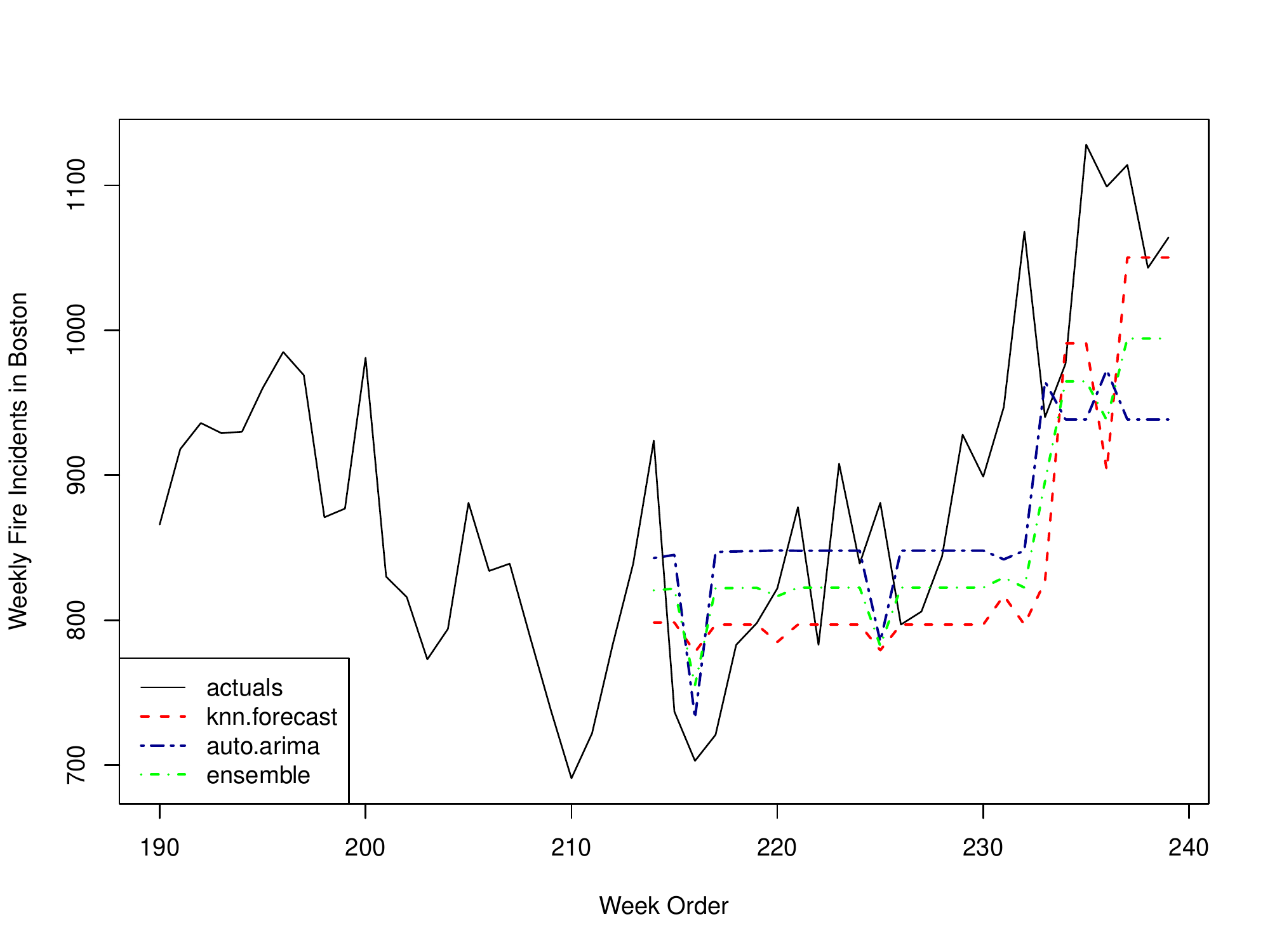} 

}

\caption{\label{fig:fire-incidents-plot} Boston fire incidents forecast comparison using KNN regression and regression with ARIMA errors.}\label{figfire-incidents-plot}
\end{figure}
\end{CodeChunk}

\section{Demonstration on simulated data} \label{section:simulatedex}

We can continue with a demonstration of \pkg{knnwtsim} on an example
using one of the simulated series included with the package. To get
started we access \code{simulation_master_list}, which was described in
Section \ref{section:data}, and select the necessary components for
forecasting. In this case using the sub-list at index 19 of the overall
list, and focus on the series which has a piece-wise functional
relationship between response and predictor as indicated in Equation
\ref{eq:quad_to_cubic}. Again as in Section \ref{section:realworldex}
storing the vector of time orders in \texttt{df\$t}, the vector of
seasonal periods in \texttt{df\$p}, the response series in \code{y} and
the exogenous predictor in \(X\), while also creating a copy of the
response series with the \code{ts} class in \code{y.ts} for use in
\code{auto.arima} from \pkg{forecast}. We then plot the full time series
of interest for reference in Figure \ref{fig:sim-data-full}. As one can
see because of the nature of the \(f(x)\) component of the series there
can be major shocks in the the response series based on the value of the
predictor variable.

\begin{CodeChunk}
\begin{CodeInput}
R> data("simulation_master_list")
R> series.index <- 19
R> y <- simulation_master_list[[series.index]]$series.quad.to.cubic.x
R> df <- data.frame(y)
R> df$t <- c(1:nrow(df))
R> p.max <- simulation_master_list[[series.index]]$seasonal.periods
R> df$p <- rep(1:p.max, length.out = nrow(df))
R> X <- as.matrix(simulation_master_list[[series.index]]$x.chng)
R> y.ts <-ts(y, frequency = p.max )
\end{CodeInput}
\end{CodeChunk}

\begin{CodeChunk}
\begin{figure}[t!]

{\centering \includegraphics{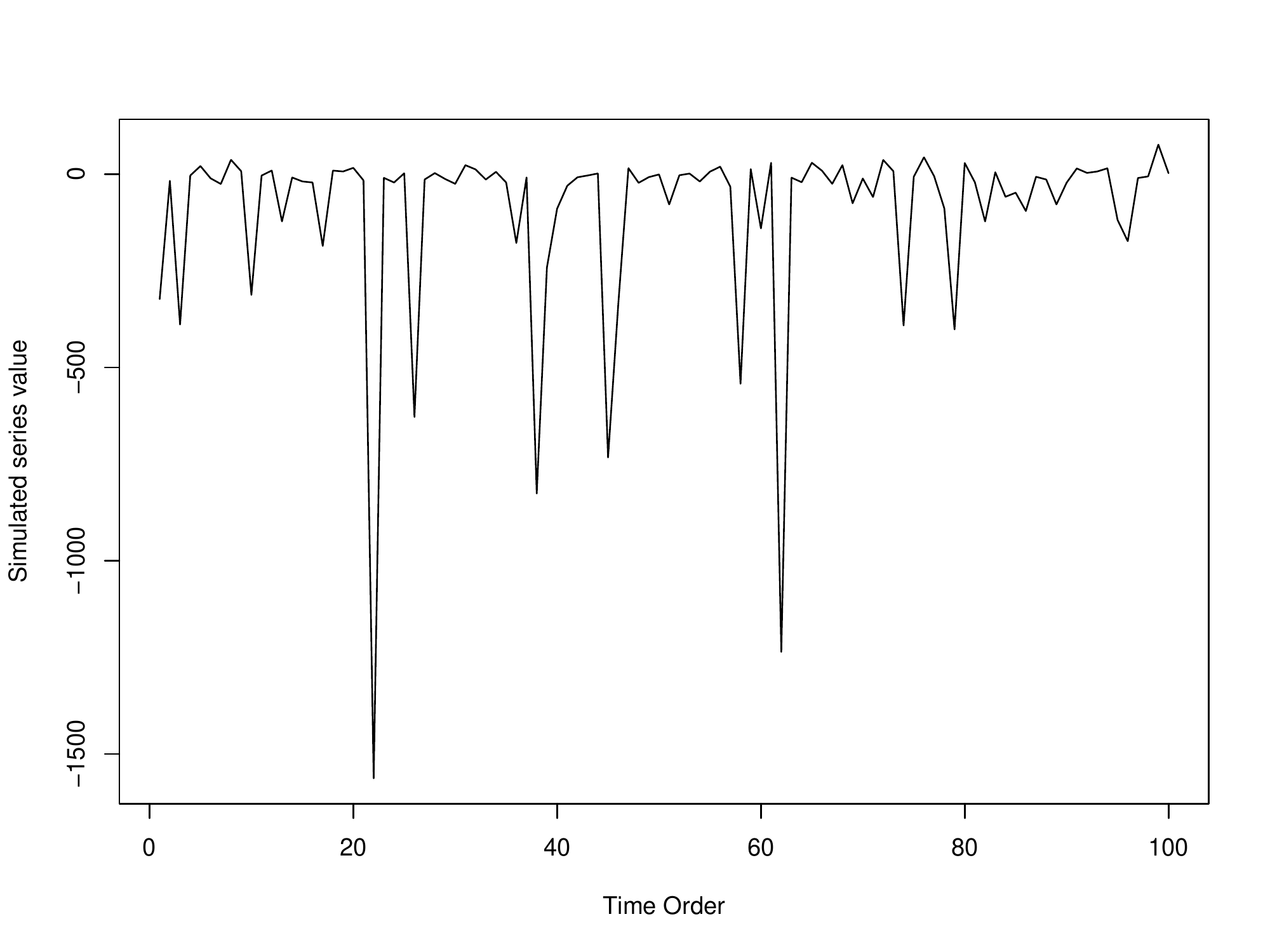} 

}

\caption{\label{fig:sim-data-full} Full simulated data series with a peice-wise functional relationship to predictor.}\label{figsim-data-full}
\end{figure}
\end{CodeChunk}

Next, in this case we can assume the hyperparameters we want to use to
calculate \code{S_w} and for the number of nearest neighbors to use in
KNN are known in order to illustrate the use of \code{SwMatrixCalc}, and
assign them to the variables \code{pre.tuned.wts} for the component
similarity weights and \code{pre.tuned.k} for the number of nearest
neighbors. To use this function we pass the time orders as the
\code{t.in} argument, the seasonal periods and the maximum seasonal
period as the \code{p.in} and \code{nPeriods.in} arguments respectively,
the exogenous predictor used in the calculation of \(S_x\) as
\code{X.in}, and finally the vector of pre-tuned weights to the
\code{weights} argument. This will call each of the component matrix
calculation functions, apply the given vector of weights, and return a
similarity matrix calculated using Equation \ref{eq:S_w}.

\begin{CodeChunk}
\begin{CodeInput}
R> pre.tuned.wts <- c(0.01350441, 0.76939463, 0.21710096)
R> pre.tuned.k <- 4
R> Sw.ts <- knnwtsim::SwMatrixCalc(
+   t.in = df$t,
+   p.in = df$p, nPeriods.in = p.max,
+   X.in = X,
+   weights = pre.tuned.wts)
R> Sw.ts[1:5, 1:5]
\end{CodeInput}
\begin{CodeOutput}
          1         2         3         4         5
1 1.0000000 0.4077682 0.3910219 0.4059744 0.7949653
2 0.4077682 1.0000000 0.4069866 0.3607408 0.4332004
3 0.3910219 0.4069866 1.0000000 0.4084143 0.2823299
4 0.4059744 0.3607408 0.4084143 1.0000000 0.4511752
5 0.7949653 0.4332004 0.2823299 0.4511752 1.0000000
\end{CodeOutput}
\end{CodeChunk}

With the similarity matrix calculated we can set the index we wish to
forecast as \code{val.index}, which in this case will be the final two
seasonal cycles each of consisting of four periods. Then call
\code{knn.forecast} to return the forecast for those points.

\begin{CodeChunk}
\begin{CodeInput}
R> val.len <- p.max * 2
R> val.index <- c((length(y) - val.len + 1):length(y))
R> final.forecast <- knnwtsim::knn.forecast(
+   Sim.Mat.in = Sw.ts,
+   f.index.in = val.index,
+   k.in = k.ts,
+   y.in = y)
R> final.forecast
\end{CodeInput}
\begin{CodeOutput}
         93          94          95          96          97 
   2.505898   -4.677692 -174.618059  -70.666159  -13.113783 
         98          99         100 
  -9.378483   -1.366484   -1.099672 
\end{CodeOutput}
\end{CodeChunk}

As in Section \ref{section:realworldex} we are interested in the
performance of this forecast relative to \code{auto.arima}, so next the
regression with ARIMA errors forecast is produced and the performance
results for both methods are captured following the approach used in
Section \ref{section:realworldex}, thus only a subset of that code is
shown below. Additionally, we forego the use of an ensemble forecast. In
terms of the reference forecast, in this case \code{auto.arima} returns
a simple linear regression with all ARIMA orders set to 0.

\begin{CodeChunk}
\begin{CodeInput}
R> y.arima.train <- y.ts[-(val.index)]
R> X.arima.train <- as.matrix(
+   X[-(val.index), ],
+   ncol = ncol(X),
+   nrow = nrow(y.arima.train))
R> arima.model <- forecast::auto.arima(
+   y.arima.train,
+   xreg = X.arima.train,
+   allowdrift = T)
R> summary(arima.model)$coef
\end{CodeInput}
\begin{CodeOutput}
intercept      xreg 
-53.99516 -24.97937 
\end{CodeOutput}
\end{CodeChunk}

Table \ref{sim-mape-table} shows the MAPE result over the final 8 points
for both methods, and Figure \ref{fig:simulated-forecast-plot} shows the
same series as Figure \ref{fig:sim-data-full} but only from time order
85 and on where the magnitude of observations are less extreme than at
certain earlier points in the series, along with the forecasts from both
methods. From both it is clear that for this forecast KNN regression is
the better performing method. In this situation the more local nature of
estimation in KNN regression is helpful, in that unless a point large in
absolute magnitude is included in the neighborhood for a point \(y_t\)
which is to be estimated, those high magnitude results will not effect
the estimate \(\hat{y}_t\). While in the case of regression with ARIMA
errors, or in this case simple linear regression alone, the estimates of
model parameters are determined more globally, based on the entire
series and so high magnitude points will influence the forecasts of all
points, even when this is not desired. While as mentioned previously
regression with ARIMA errors is often an excellent solution to
univariate forecasting, this series presents a case where a
nonparametric approach like KNN regression with \(S_w\) may be
preferred.

\begin{table}[t!]
\centering
\begin{tabular}{rrr}
  \hline
 & KNN regression & Regression with ARIMA errors \\ 
  \hline
1 & 79.07 & 1298.65 \\ 
   \hline
\end{tabular}
\caption{Simulated data MAPE results by forecast method.} 
\label{sim-mape-table}
\end{table}

\begin{CodeChunk}
\begin{figure}[t!]

{\centering \includegraphics{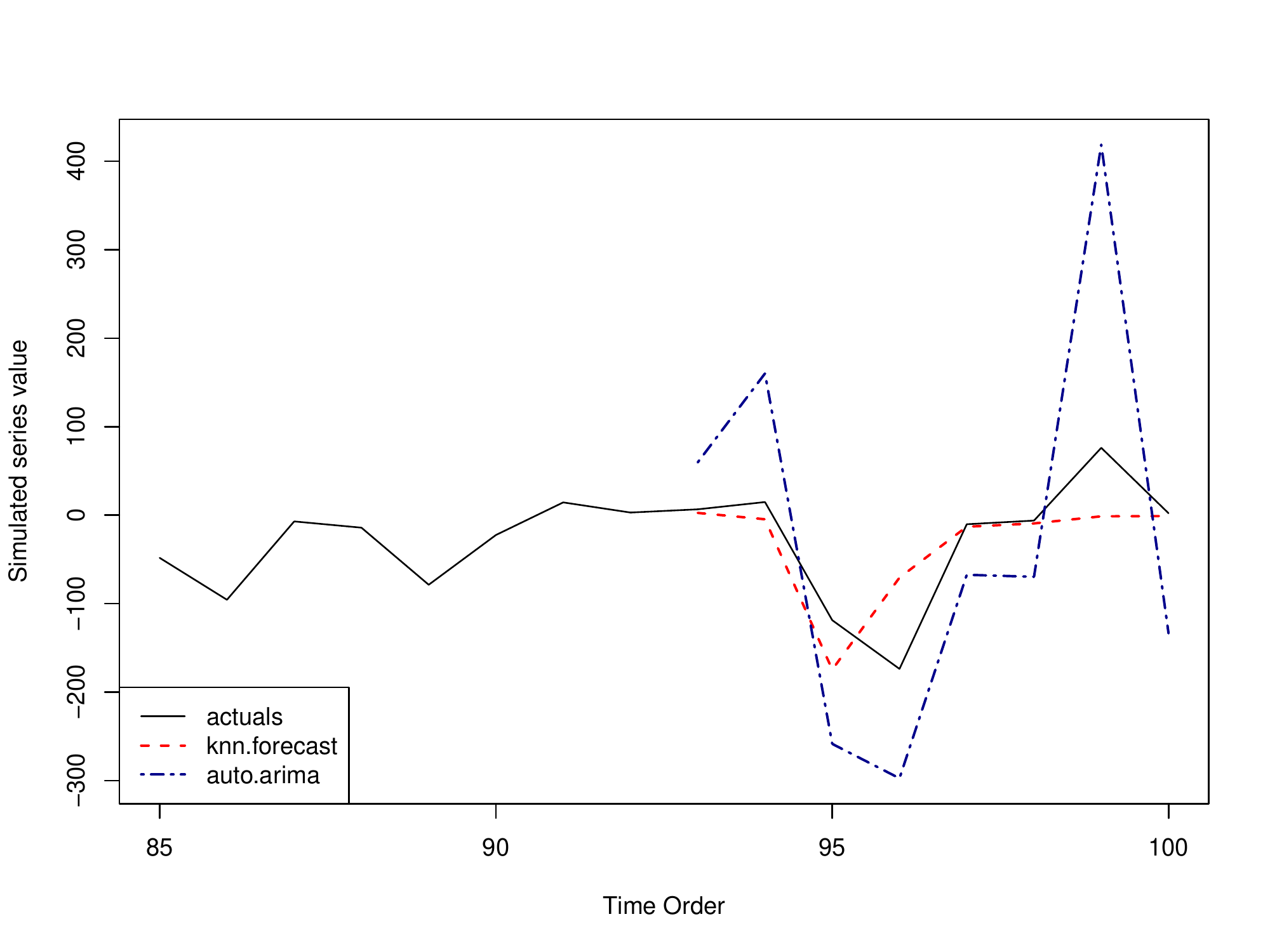} 

}

\caption{\label{fig:simulated-forecast-plot} Simulated data forecast comparison using KNN regression and regression with ARIMA errors}\label{figsimulated-forecast-plot}
\end{figure}
\end{CodeChunk}

\section{Conclusion} \label{section:conclusion}

The \pkg{knnwtsim} package provides an alternative to existing software
for those interested in univariate forecasting with KNN regression. In
this article the similarity metric \(S_w\) was introduced as well as all
of of its components, which could themselves be used alone in KNN
regression if desired, as was KNN regression based forecasting more
generally as implemented in \pkg{knnwtsim} in Sections
\ref{section:simform} and \ref{section:knnforecasting} respectively.
Additionally, the functions and data contained in \pkg{knnwtsim} are
detailed in Section \ref{section:package}, and demonstrated through
examples in Sections \ref{section:realworldex} and
\ref{section:simulatedex}.

\section{Acknowledgements} \label{section:acknowledgements}

I would like to thank Ernest Fokoue for his contribution to this package
and article, he has been an immensely helpful advisor throughout the
development of \pkg{knnwtsim} and there would not be a package without
his advice and support.

\bibliography{references.bib}

\end{document}